\documentclass[letter]{aa}  

\usepackage{natbib,graphicx,amssymb,amsmath}
\bibpunct{(}{)}{;}{a}{}{,} 
\newcommand{\kms}{$\rm km\,s^{-1}$}

\begin{document}
\title{Chemistry in isolation: High CCH/HCO$^+$ line ratio in the AMIGA galaxy CIG~638}
\titlerunning{Chemistry in isolation: CIG~638}
\author{
  S. Mart\'in\inst{\ref{inst1}}
  \and L. Verdes-Montenegro\inst{\ref{inst2}}
  \and R. Aladro\inst{\ref{inst3}}
  \and D. Espada\inst{\ref{inst4},\ref{inst5},\ref{inst6}}
  \and M. Argudo-Fern\'andez\inst{\ref{inst2}}
  \and C. Kramer\inst{\ref{inst7}}
  \and T.C. Scott\inst{\ref{inst2}}
}
\institute{
  Institut de Radio Astronomie Millim\'etrique, 300 rue de la Piscine, Dom. Univ., 38406, St. Martin d'H\`eres, France\\
  \email{smartin@iram.fr}\label{inst1}
  \and
  Instituto de Astrof\'isica de Andaluc\'ia, Granada, IAA-CSIC Apdo. 3004, E-18080 Granada, Spain\label{inst2}
  \and
  European Southern Observatory, Alonso de C\'ordova 3107, Vitacura, Casilla 19001, Santiago 19, Chile\label{inst3}
  \and
  Joint ALMA Observatory (JAO), Alonso de C\'ordova 3107, Vitacura, Santiago, Chile\label{inst4}
  \and
  National Astronomical Observatory of Japan (NAOJ), 2-21-1 Osawa, Mitaka, 181-8588, Tokyo, Japan\label{inst5}
  \and
  Department of Astronomical Science, The Graduate University for Advanced Studies (SOKENDAI), 2-21-1 Osawa, Mitaka, 181-8588, Tokyo, Japan\label{inst6}
  \and
  Instituto Radioastronom\'ia Milim\'etrica, Av. Divina Pastora 7, E-18012 Granada, Spain\label{inst7}
}

\abstract
{
Multi-molecule observations towards an increasing variety of galaxies have been showing that the relative molecular abundances are affected by the type of activity.
However, these studies are biased towards bright active galaxies, which are typically in interaction.
}
{
We study the molecular composition of one of the most isolated galaxies in the local Universe where the physical and chemical properties of their molecular
clouds have been determined by intrinsic mechanisms.
}
{
We present 3~mm broad band observations of the galaxy CIG~638, extracted from the AMIGA sample of isolated galaxies. The emission of the $J=1-0$
transitions of CCH, HCN, HCO$^+$, and HNC are detected. Integrated intensity ratios between these line are compared with similar observations
from the literature towards active galaxies including starburst galaxies (SB), active galactic nuclei (AGN), luminous infrared galaxies (LIRG), and GMCs in M33.
}
{
A significantly high ratio of CCH with respect to HCN, HCO$^+$, and HNC is found towards CIG~638 when compared with all other galaxies where these species have
been detected. This points to either an overabundance of CCH or to a relative lack of dense molecular gas as supported by the low HCN/CO ratio, or both.
}
{
The data suggest that the CIG~638 is naturally a less perturbed galaxy where a lower fraction of dense molecular gas, as well as a more even distribution
could explain the measured ratios. In this scenario the dense gas tracers would be naturally dimmer, while the UV enhanced CCH, would be overproduced
in  a less shielded medium.
}

\maketitle
\section{Introduction}
Deep systematic multi-molecular studies in galaxies have been carried out for almost a decade now and have been significantly boosted by the recent
increase in the instantaneous bandwidth of mm and submm facilities \citep[see][for a review]{Mart'in2011a}. Both targeted and unbiased surveys
have been carried out towards the nuclei of the brightest and nearest prototypical active galaxies 
\citep{Wang2004,Mart'in2006,Mart'in2011,Aladro2011a,Aladro2013} with the aim of studying the effects of the nuclear activity on the overall molecular abundances.
Studies using small samples 
of galaxies have resulted in some potential activity diagnostics based on molecular abundance ratios
\citep{Kohno2001,Mart'in2009,Costagliola2011}.
In particular, the ratio between HCN and HCO$^+$ has been proposed as a potential discriminator of the activity type whether AGN or SB dominated
\citep{Kohno2001,Krips2008}. In this scenario, the HCN abundance would be enhanced by the pervading X-ray radiation from the
central black hole, while the HCO$^+$ would be favoured in UV irradiated and/or cosmic ray enhanced star-forming regions.

However, all these studies have been naturally limited to the brightest, more active nearby galaxies.
The AMIGA isolated galaxy sample \citep{Verdes-Montenegro2005,Verdes-Montenegro2010} has clearly established that parameters expected to
be enhanced by interactions, such as $L_{FIR}$, radio continuum emission, AGN rate, or HI asymmetry, are lower in isolated galaxies than in any other sample,
even compared with field galaxies.
In particular, the comparison between the AMIGA sample and a sample of Hickson compact groups found indications that
molecular gas in isolated galaxies could be more extended than in environments with a higher density of galaxies \citep{Lisenfeld2011,Martinez-Badenes2012}.
In this paper, we aim to explore the chemistry of the most isolated galaxies in the local Universe in order to add a new type of galaxy to these
comparative molecular studies.
It is expected that interactions will have a direct effect on the physical conditions of giant molecular clouds
(GMCs), the star formation history, and activity within the nuclear regions of galaxies, which will very likely lead
to a different chemical evolution of the molecular material in the galaxy.
Interestingly, only eight LIRGs and no ULIRG are found in the AMIGA sample. 
A comparison based on molecular abundances can provide hints to how the secular evolution affects the ISM associated to
a nuclear activity not triggered and/or affected by interactions.
Therefore, galaxies in isolation might be key as chemical baselines to understand the ways in which the ISM has evolved in isolation as opposed to other more active
interacting galaxies.

\begin{figure}
\centering
\includegraphics[width=\linewidth]{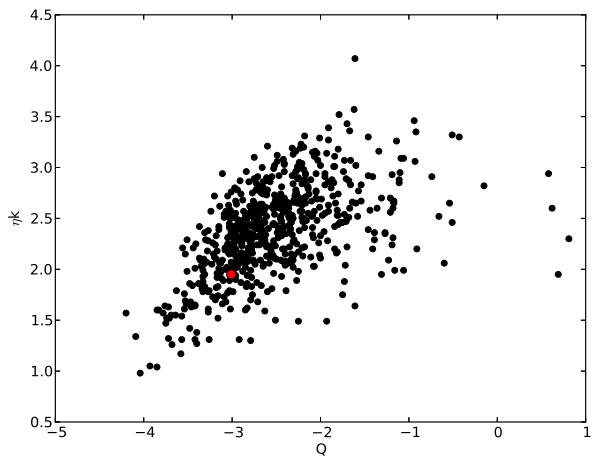}
\caption{Location of the number density and tidal parameters of CIG~638, as obtained from the isolation parameters in \citet{Argudo-Fernandez2013}, compared to the whole AMIGA sample. 
 \label{fig.isolation}}
\end{figure}

The star-forming galaxy CIG~638 (NGC~5690)
matches the isolation criteria from \citet{Verley2007b}  
based on revised local number density and tidal parameters (Fig.~\ref{fig.isolation}) from \citet{Argudo-Fernandez2013}.
A revision of isolation using spectroscopic data shows CIG~638 is isolated from similar luminosity neighbours, unlike the comparison
galaxies used in this paper, for which data exist for evaluating isolation.
Also, in the $L_{\rm FIR}-L_{\rm B}$ and $M_{\rm H_2}$ vs $L_{\rm B}$ plots, CIG~638 lies in the same parameter space as the bulk of the AMIGA sample.
Its morphological type (Sbc) is the most abundant in AMIGA, and the symmetric HI profile also reinforces its isolation \citep{Espada2011}.
While the degree of isolation of nearby galaxies ($v<1500$~\kms) cannot be reliably determined, the more distant galaxies, being quiescent, tend to be fainter at all wavelengths.
In fact, the vast majority of isolated galaxies in the sample are not bright enough in CO \citep{Lisenfeld2011} to easily detect other molecular species than CO,
typically a factor of 20 to 100 fainter.
The bright ($\sim50$~mK) and broad ($FWZI\sim280$~\kms) CO emission of the almost edge-on CIG~638 ($i=78^\circ$) makes it the ideal CO-bright
isolated galaxy for deep molecular observations.

\begin{figure*}
\centering
\includegraphics[width=0.9\linewidth]{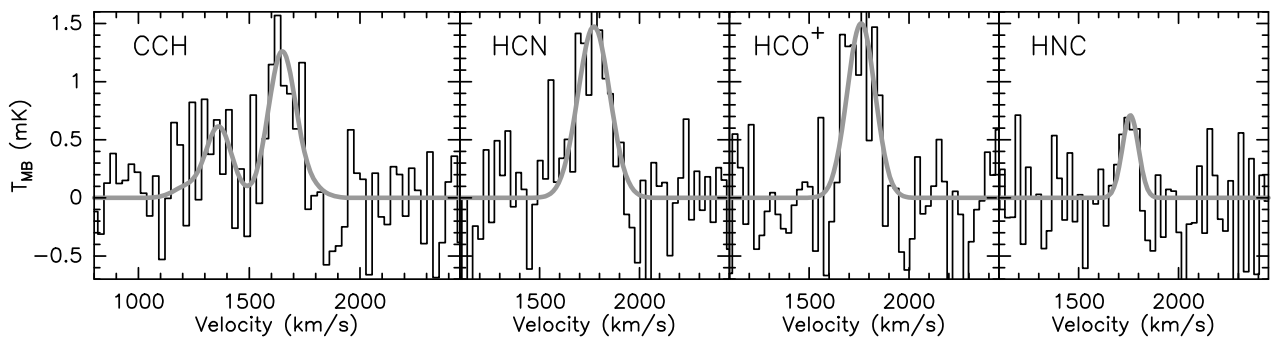}
\caption{
Detected $J=1-0$ transitions towards CIG~638. Spectral resolution is smoothed down to 26~\kms.
 \label{fig.spectra}}
\end{figure*}

\section{Observations}
Observations were carried out with the IRAM~30m telescope in Pico Veleta, Spain, in two different runs in 
November 2011 and October 2012. The E090 receivers were tuned to cover $2\times8$~GHz bands centred at $\sim89.6$~GHz and $\sim105.4$~GHz in dual
polarization. The spectral resolution of 200~kHz from the FTS backend was smoothed down to a coarser resolution of
7.8~MHz (26~\kms~ at $\sim90$~GHz), suitable for the broad observed emission.
Symmetrical wobbler-switching mode observations were performed with
an off position $220''$ away from the source in azimuth.
Pointing and focus were regularly checked towards bright nearby continuum quasars.
Spectra were calibrated using the standard dual load system. Typical system temperatures of $T_{sys}=90-130$~K were measured.
We achieved an rms of 0.3-0.4~mK at 26~\kms~ resolution in a total integration time of 22.13 hours.
The nominal observing position was $\alpha_{J2000}=14^h37^m41\fs10,~\delta_{J2000}=02^\circ17'27.0''$ where,
at a distance of 19.7~Mpc \citep{FernandezLorenzo2012}, the $28''$ beam covered the central 2.6~kpc of CIG~638.


\section{Results}
We detected the $J=1-0$ transitions of CCH, HCN, HCO$^+$, and HNC (Fig.~\ref{fig.spectra}).
Synthetic spectra were fitted to the observations using the SLIM package within MADCUBA\_IJ (Mart\'in et al. in prep.) 
where column density, velocity, and width parameters are fit.
Table~\ref{tab.params} presents derived parameters from best fit in Fig.~\ref{fig.spectra}
assuming an excitation temperature of $T_{ex}=15$~K and source size of $5''$
($\sim500$~pc), a reasonable size for the central molecular region as observed in nearby galaxies and our own.
These assumptions have an impact on the absolute column densities, but the line ratios discussed
in this paper are not affected.
Moreover, simultaneous observations with wideband receivers 
allow an accurate line intensity ratio determination.
The largest uncertainty results from the assumption of all species having the same distribution that, as shown by high-resolution observations \citep{Meier2005},
is likely not to be true.

\begin{table}
\caption{Fit parameters to the observed lines\label{tab.params}}
\centering
\begin{tabular}{l@{\,\,\,}l@{\,\,\,}l@{\,\,\,}l@{\,\,\,}l@{\,\,\,}l}
\hline\hline
Molec.              & $\nu$   & $N$                            & $\Delta v_{1/2}$  & $T_{\rm MB}$  &  $\int{T_{\rm MB}{\rm d}v}$   \\ 
                    & (GHz)   & $(\times10^{13}\,\rm cm^{-2})$ &  (km\,s$^{-1}$)   & (mK)          &  (mK\,km\,s$^{-1}$)           \\ 
\hline
CCH\tablefootmark{a}& 87.317  &   33 (5)                       & 130 (30)         & 0.9        &  300 (40)                         \\ 
HCN                 & 88.630  &  2.1 (0.3)                     & 185 (30)         & 1.5        &   290 (40)                        \\ 
HCO$^+$             & 89.188  &  1.10 (0.18)                   & 163 (30)         & 1.5        &   260 (40)                        \\ 
HNC                 & 90.663  &  0.47 (0.12)                   & 90 (30)          & 0.7        &   70 (20)                         \\ 
\hline
\end{tabular}
\tablefoot{
All lines were centred at $1770\pm15$~\kms.
\tablefoottext{a}{Integrated intensity refers to the whole CCH hyperfine group, while the other parameters refer to the brightest component.}
}
\end{table}

To understand the peculiarities of the molecular abundances in CIG~638, we looked up for a sample of comparison sources with
available simultaneous line observations.
We collected 23 observational data of the same four molecular transitions towards 19 sources from \citet{Costagliola2011},\citet{Jiang2011}, \citet{Aladro2013}, and \citet{Davis2013}.
The sample includes starburst galaxies (SB), active galactic nuclei (AGN), and luminous infrared galaxies (LIRGs)
which are very likely a composite of SB and AGN.
We added the stacked spectrum towards seven GMCs across the metal-poor spiral galaxy M33 to this sample \citep{Buchbender2013}.
In the work by \citet{Costagliola2011}, the hyperfine structure of CCH has been fit by a single Gaussian. 
Using the synthetic spectra provided by MADCUBA\_IJ, one can show that for sources with broad spectral lines $200-500$~\kms wide, 
the CCH line will appear $\sim200$\~kms~ owing its hyperfine splitting. 
Therefore we consider outliers any source by \citet{Costagliola2011} with CCH fit width
$>300$\kms~larger than that of HCN, namely
IC860, NGC3556, and UGC5101.

\subsection{Line ratio comparison}
\begin{figure}
\centering
\includegraphics[width=0.9\linewidth]{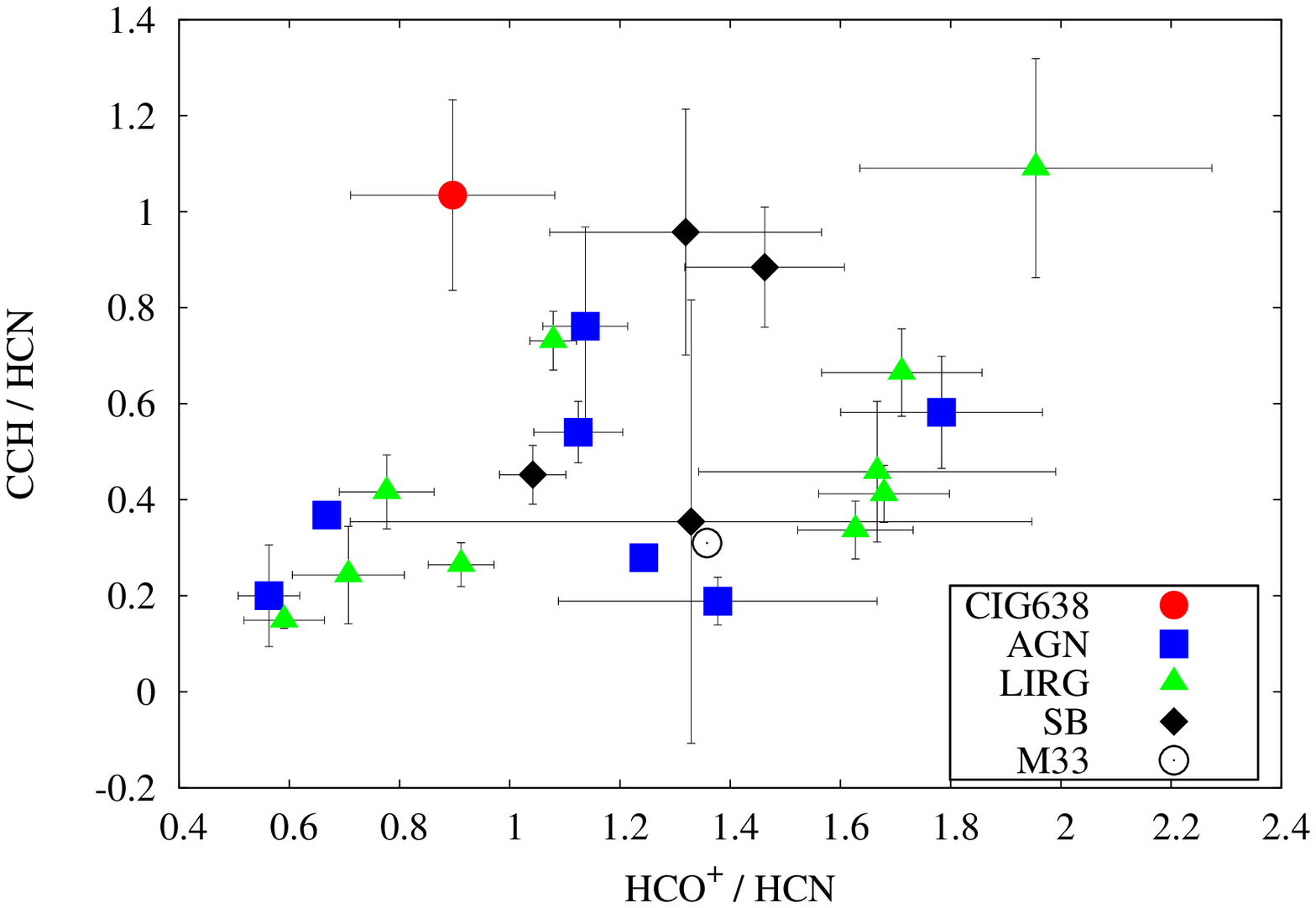}
\includegraphics[width=0.9\linewidth]{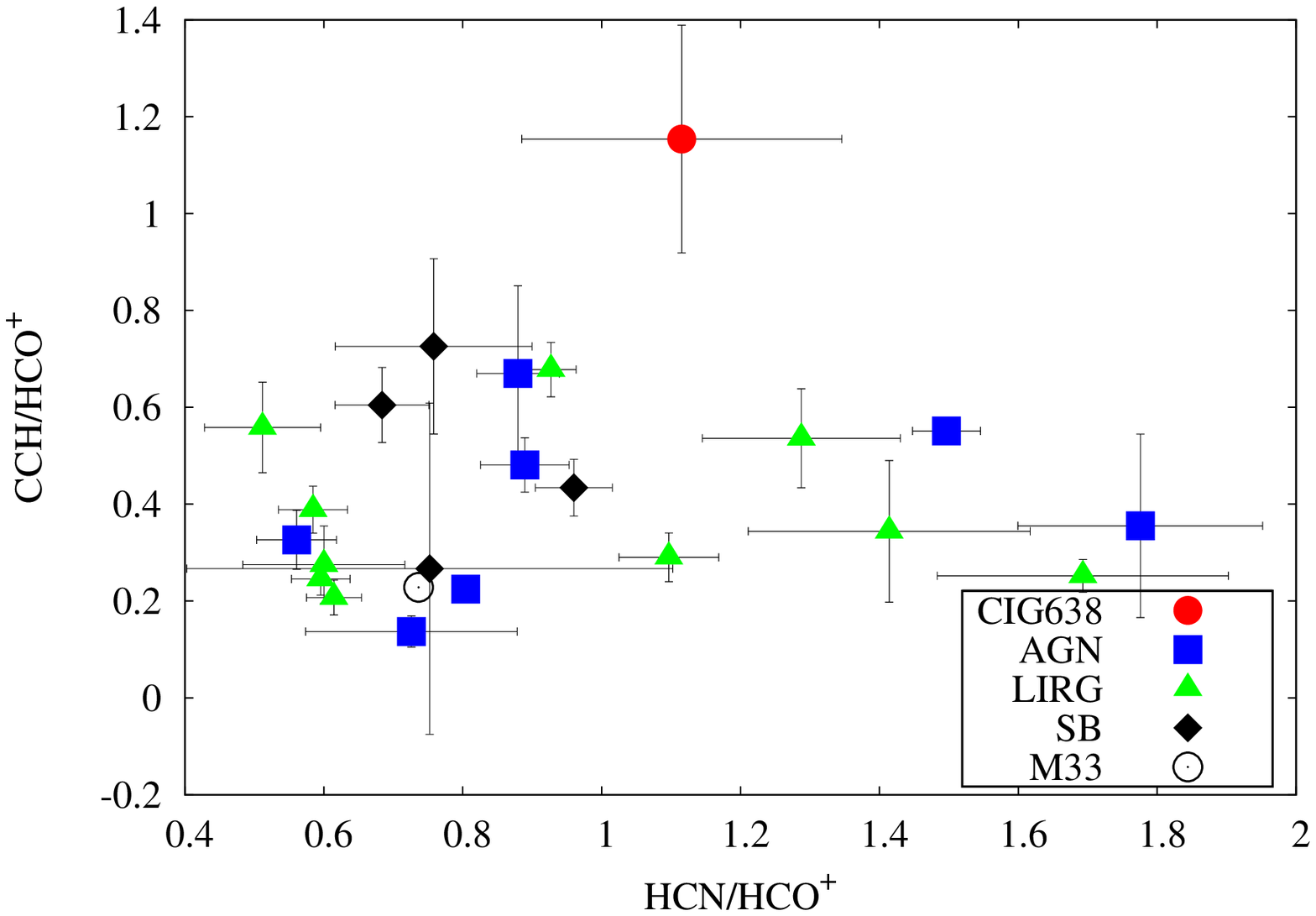}
\includegraphics[width=0.9\linewidth]{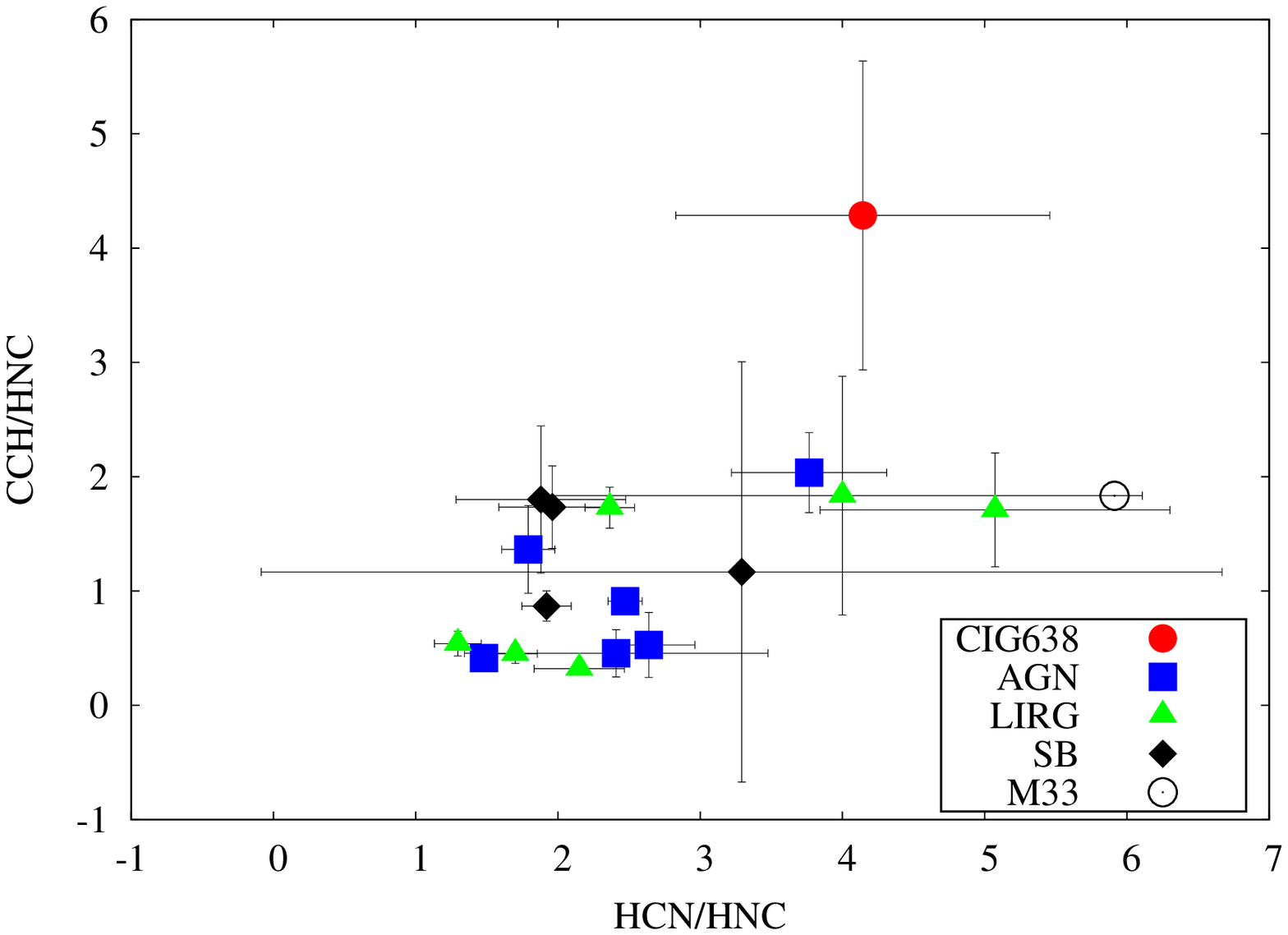}
\caption{CCH, HCN, HCO$^+$ and HNC integrated intensity ratios.
 \label{fig.comparison}}
\end{figure}

The comparison of the relative integrated intensity of CCH to those of HCN, HCO$^+$, and HNC against the line ratios between HCN, HCO$^+$ and HNC are
shown in Fig.~\ref{fig.comparison}.
The HCN/HCO$^+$ ratio 
of the galaxy comparison sample ranges the values $0.5-2.4$
showing even larger variations than those measured by \citet{Krips2008}. 
While the SB galaxies show values $>1$, AGN are vary from very low ratios to values greater than unity in which the starburst contribution
may be significant. LIRGs, on the other hand, are found all over the parameter space due to their diverse nuclear activities.
CIG~638 is observed to have a ratio of HCN/HCO$^+=0.90\pm0.18$, which is slightly below the average of the
observed distribution of ratios in the galaxy sample. 
Thus, from this observed ratio, neither of these species shows an enhancement which could be attributed to a particular
heating mechanism in the ISM of CIG~638.

The CCH/HCN ratio 
shows a significant variation ranging from 0.15 to 1.1 where we find CIG~638 to be among the
highest values. The only galaxy showing a higher intensity ratio
is the LIRG NGC~3690, while the SB galaxies NGC~4194 and UGC2866 show similar ratios.

The left-hand panel in Fig.~\ref{fig.comparison} also shows an increasing trend that, since it is normalized by HCN in both axes, 
is the result of the correlation between HCO$^+$ and CCH.
In fact, the middle panel in Fig.~\ref{fig.comparison} shows how the CCH/HCO$^+$ ratio has a relatively smaller variation between galaxies and is more 
concentrated around $0.4\pm0.3$. Thus, it appears to be a loose correlation between these species.
The line ratio observed in CIG~638 is clearly above the one observed in any other galaxy, which points towards a real abundance differentiation with
respect to other active galaxies.

Similarly, CIG~638 shows a CCH/HNC ratio that is a factor of 2 to 4 higher than any other galaxy. However, this ratio is likely to be biased by the low
signal-to-noise of the HNC detection. Still, even taking the errors into account, it appears that the CCH relative abundance is at least on the higher
end of the range observed towards other galaxies.

\section{Discussion: The abundances in CIG~638}
The previous comparison shows that the integrated intensity of CCH relative to any other of the observed species appears to be significantly
higher than in any other galaxy in our bibliographic sample. These observations clearly show a difference in the molecular abundances of CIG~638, which
could either be the result of an overproduction of CCH or an underabundance of HCN, HCO$^+$, and HNC.

We first explore the latter scenario.
Given that all ratios between the species HCN, HCO$^+$, and HNC appear to agree with those in the comparison sample, one would need all these
species to be underabundant with respect to CCH in CIG~638 to explain their measured ratios.
In fact, all comparison sources from the literature are active galaxies where we expect large fractions of dense molecular gas associated to the
bursts of star formation, fueling the central black hole or, more generally by compressing material from interaction events.
Though containing significant amounts of dense molecular gas to maintain its observed star formation, which is responsible
for its $L_{FIR}=5\times10^9~L_\sun$, the galaxy might be deficient in dense molecular gas as traced by HCN.
If we calculate the integrated intensity ratio between the $1-0$ transitions of CO and HCN as a proxy for such overall to dense molecular gas fraction,
we obtain a range of $100-150$ for CIG~638 which is significantly higher than the ratios of $\sim20-50$ calculated from the sample from \citet{Costagliola2011} and $\sim77$ for 
the stacked GMCs in M33. 
Though this is just a rough estimate, the significantly larger CO/HCN ratio suggest that CIG~638 might be deficient in dense molecular gas.
Interestingly, the HCN/CO$=1-0.6\%$ and HCO$^+$/CO=$0.9-0.5\%$ ratios in CIG~638 is comparable to three of the GMCs in M33 which, though massive, show the lowest star formation
efficiency \citep{Buchbender2013}. The low ratios for these GMCs in M33 are claimed to indicate a low dense gas fraction.
The CO observations used to estimate this ratio in CIG~638 were carried out at the 14~m FRAO telescope by \citet{Lisenfeld2011}.
In order to estimate the expected intensity within the IRAM 30m beam, we corrected by the different beam dilution for a conservative range of emission extents of
$1''(100pc)<\theta_s<20''(2Kpc)$.
Only when CO emission is much more extended, the ratio would be similar to those in the sample from \citet{Costagliola2011}.
That, however, would support the idea of most of the molecular gas CIG~638 being rather diffuse and only a minor fraction in the form of dense GMCs.

A second scenario implies an overproduction of CCH in CIG~638.
Since the first extragalactic detection by \citet{Henkel1988}, not much attention has been paid to the CCH molecule outside the Galaxy.
The study by \citet{Nakajima2011} towards the SB NGC~253 and the AGN NGC~1068 suggests that the nuclear activity had no observable
effect on the CCH abundance relative to CS, as supported by the modelling by \citet{Aladro2013}. Similarly, our measured CCH relative abundances show 
no trend with the nuclear activity, supporting this idea.
\citet{Jiang2011} observed a tentative CCH/HCN and CCH/HCO$^+$ decrease with increasing $L_{FIR}$ that could suggest that the overabundance in
CIG~638 might be related to its relatively low luminosity. However, several of the sources in our comparison sample have similar luminosities to CIG~638
but do not show such overabundance.

Within the Galaxy, CCH observations show that the abundance of this molecule is increased in the early stages of star formation.
CCH is subsequently depleted from the cores as the star formation proceeds, but the higher abundances of CCH are then displaced towards the outer
cloud edges that are highly pervaded by interstellar UV photons \citep{Beuther2008,Walsh2010,Li2012}.
This is indeed observed in external galaxies where CCH is prevailing in highly UV radiated regions \citep{Meier2005}.
CIG~638 is unlikely to be dominated by PDRs that are the result of past starburst events. However, the presence of a widespread, less dense molecular
material or fewer high density clouds would result in an overall lower visual extinction, allowing the pervading UV radiation to
inflict a chemical footprint on the CCH abundance.

\section{Conclusions}
Based on single-dish line ratio observations of the isolated galaxy CIG~638, molecular abundance differences are found when compared to
local interacting active galaxies.
The lower tidal force in CIG~638 than in our comparison sample has an influece not only on the nuclear activity
\citep{Sabater2013} but also on the overall ISM physical conditions.
Assuming that CIG~638 is representative of isolated galaxies, 
we suggest that in isolation a lower fraction of dense gas and a more homogeneously distributed, low-extinction ISM 
leads to both a low abundance of dense gas tracers, such as HCN and HCO$^+$, and an overabundance of CCH due to poorer shielding from UV
radiation.
However, spatially resolved sensitive imaging on a larger sample of isolated galaxies is required to confirm this scenario.

\begin{acknowledgements}
Based on observations carried out with the IRAM 30m telescope. IRAM is supported by INSU/CNRS (France), MPG (Germany), and IGN (Spain).
We thank Mirian Fern\'andez Lorenzo for valuable discussion of the isolation of CIG~638.
This work has been supported by grant AYA2011-30491-C02-01 co-financed by MICINN and FEDER funds, and the Junta de Andalucia (Spain) grants P08-FQM-4205 and TIC-114.
\end{acknowledgements}

\bibliographystyle{aa}	
\bibliography{CIG638.bib}	

\end{document}